\documentclass[prd,aps,showpacs,tightenlines,superscriptaddress,amsmath,amssymb,amsfonts,twocolumn]{revtex4}
\usepackage{amssymb,amsmath,amsthm,amsbsy,epsfig,color,graphicx,times,bbm}
\usepackage[ansinew]{inputenc}
\usepackage[english]{babel}
\usepackage{accents}
\usepackage{color}
\usepackage{braket}
\usepackage{booktabs}
\usepackage{tabularx}
\usepackage{array}
\usepackage{bbold}

\begin{document}

\title{Testing  CPT violation, entanglement and gravitational interactions in particle mixing with trapped ions}

\author{Antonio Capolupo}
\affiliation{Dipartimento di Fisica ``E.R. Caianiello'' Universit\'a di Salerno, and INFN - Gruppo Collegato di Salerno,  Via Giovanni Paolo II, 132, 84084 Fisciano, Italia}
\author{Salvatore Marco Giampaolo}
\affiliation{Ru\dj{}er Bo\v{s}kovi\'{c} Institute, Bijen\u{c}ka cesta 54, 10000 Zagreb, Croatia}
\author{Aniello Quaranta}
\affiliation{Dipartimento di Fisica ``E.R. Caianiello'' Universit\'a di Salerno, and INFN - Gruppo Collegato di Salerno,  Via Giovanni Paolo II, 132, 84084 Fisciano, Italia}


\begin{abstract}
By analyzing the analogies between the effective system of $N$ spins described by the Ising Hamiltonian
and the phenomenon of the self--gravity in mixed particle systems, we show that cooled ions held in a segmented ion trap
and exposed to a magnetic field gradient can simulate the proposed mechanism of mutual interaction in mixed neutrino system.
We show that with trapped ions one can reproduce the expected corrections to the flavor transitions and the
$CPT$ violation induced by gravity on flavor fields,  which may have played an important role in the early stages of the universe.
The results presented are experimentally testable. They indicate that ions confined in microtraps can represent a new tool to test fundamental phenomena of nature.
\end{abstract}

\maketitle

\section{introduction}

Particle mixing and oscillations represent the most compelling indication of physics beyond the standard model of particles~\cite{Wu1957,Giunti2007,Griffiths2008}.
The physical oscillating fields, named flavor fields, are superpositions of free fields  with different definite masses.
Particle mixing interests axion--photon system ~\cite{Raffelt1996,Sikivie2011,Capolupo2019_1}, $\eta$--$\eta'$~\cite{Pham2015}, neutral kaons~\cite{Christenson1964}, $B^{0}$ and $D^{0}$  mesons~\cite{Abashian2001}, in boson sector;  neutrino flavor oscillations~\cite{Bilenky1978,Gonzalez2008},   neutron--antineutron oscillations  ~\cite{Phillips2016}, and the quark mixing~\cite{Nakamura2010}, in fermion sector.
Many efforts have been devoted to understanding the origin of the phenomenon and, recently, also the effects of gravity on neutral mixed particles have been analyzed ~\cite{Bose2017,Marletto2017,Marletto2018,Kyrylo2019}.
Moreover, gravity has been also considered as a source of decoherence in flavor oscillations~\cite{Ellis1984,Amelino-Camelia2000,Rovelli2004}
and the effects of decoherence for mixed particles on fundamental symmetries of nature have been recently studied  ~\cite{Gago2001,Guzzo2016,Lisi2000,Benatti2000,Benatti2001,Capolupo2018,Capolupo2019,Buoninfante2020}.
It has been shown that gravity can affect the frequency of oscillations~\cite{Marletto2018,Kyrylo2019} and leads to many interesting effects like the CPT-symmetry violation in particle mixing~\cite{Gago2001,Guzzo2016,Lisi2000,Benatti2000,Benatti2001,
Capolupo2018,Capolupo2019}. Effects of quantum field theory in curved space--time on mixed neutrinos have been also analyzed \cite{CapolupoCurv2020}.
The phenomena mentioned above are difficult to reveal experimentally, due to the elusive nature of neutrinos and the short lifetime of composite particles (as pertaining to meson mixing).

Another particularly thriving field of research is represented by the study of trapped ions.
Trapped ions are among the most promising systems for practical quantum computing and
 many quantum technological applications have been developed by using  such systems.
An ion trap is a combination of electric or magnetic fields used to capture ions realizing a system well isolated from the external environment. Therefore it is characterized by long lifetimes and coherence~\cite{Trap-a,Trap-b}, strong ion-ion interactions~\cite{Trap1}, and the existence of cycling transitions between internal states of ions that allows to realize both measurement and laser cooling.
Trapped ions, besides being  one of the
leading technology platforms for large-scale quantum computing, allow the simulation of quantum models of strongly interacting
quantum matter and enable the investigation of those quantum dynamics that remain so far unexplored due their inescapable complexity~\cite{Trap-c,Trap-d}.

In this work, we show that trapped ions can also simulate the gravity effect on mixed neutral particles and thus reproduce, in table top experiments, analogues of phenomena that may have characterized the first stages of the evolution of the Universe.

In particular, we analyze the analogies between a self-interacting system of mixing particles and an ensemble of Doppler cooled ions loaded in a magnetic trap.
We note that an effective spin-$1/2$ long range Ising model can describe both Doppler cooled ions held in a segmented ion trap~\cite{Trap1,Trap2,Trap3,Trap4,Trap5}
and the dynamics of a gravitationally  self-interacting system made of oscillating neutrinos, as introduced in Ref.~\cite{Kyrylo2019}.
In that paper, it has been studied an ensemble of $N$ mixed self--interacting neutral particles as a closed system and it has been shown that  the self--gravity effects induce  a  $CPT$ violation and a modification of the oscillation formulae for systems like neutrinos.
The emerging entanglement and the $CPT$ violation are extremely small for systems like neutrinos or neutrons,
yet such effects might play a crucial role in dense matter and during the early stages of the universe.

Here, we consider different interaction potentials and we analyze the oscillation formulae and the $T$-symmetry violation for traps loaded with $4$ or $6$ ions.
We start by analyzing the case with $4$ ions and coupling in three independent wells, in which the inner ions are confined in a common well and coupled strongly.
Then we consider $6$ ions confined in a single strongly an-harmonic well.
We show that trapped ions  reproduce the expected corrections to the flavor
transitions and the $CPT$ violation induced by gravity on flavor fields.
We show that
 the $T$ violation for confined ions is the analogue of the $CPT$ violation for flavor mixed fields and
we find that $T$ violation depends on the number of trapped ions, as predicted for $CPT$ violation in the flavor mixing case.
Our numerical results show that such violations, together with the new oscillation formulae are all detectable in the present experiments on ion traps.
Reproducing the effects predicted for neutrinos in atomic systems could open new scenarios for the study of fundamental physics; it could be of high relevance in the understanding of the evolution of the universe and possibly lead to the discovery of new phenomena.

The paper is organized as follows.
In Sec.~II, for the reader's convenience, we briefly resume the main results of the treatment of gravitationally self-interacting oscillating neutrinos.
In Sec.~III, we present the analogies with cooled trapped ions and we show that the very tiny effects induced by the gravitational self-interaction for neutrinos have corresponding, larger effects in trapped ions, which can be easily detected.
Indeed, we show that the oscillation formulae for trapped ions and the $T$ symmetry violation which can be interpreted as a signature of $CPT$ violation in mixed flavor fields, can be experimentally tested with the present technologies.
In Sec.~VI we draw our conclusions.

\section{Oscillations of N interacting particles}

In this section we report the main results obtained for oscillating and gravitationally interacting particles~\cite{Kyrylo2019}.
The mixing relations for two flavor fields $ n_A $ and $ n_B $ are
\begin{eqnarray}
\label{leptonic_state}
\ket{n_A} & = & \cos(\theta) \ket{m_1}+ e^{\imath \phi } \sin(\theta) \ket{m_2} \ \nonumber \,;\\
\ket{n_B} & = & - e^{-\imath \phi } \sin(\theta) \ket{m_1}+ \cos(\theta) \ket{m_2} \, ,
\end{eqnarray}
where $ n_A = \nu_e $ and $ n_B = \nu_\mu $ for neutrinos, $\theta$ is the mixing angle, $\phi$ is the Majorana phase which is zero in the case of Dirac fermions~\cite{Giunti2007} and $\ket{m_i}$ are the states with definite masses $m_i$.

Denoting with $E$ the energy of the mixed particles travelling through the space, and assuming $m_i \ll E$, the Hamiltonian of mixed fields can be written as
\begin{eqnarray} \label{Hamiltoniana_singolo_2}
H =\omega_0 \, \sigma^z \; ; \;\;\;\;\;\;\;\;\; \omega_0=\frac{c^2}{4 E}(m_1^2-m_2^2) \,,
\end{eqnarray}
where $\sigma^z\!=\!\ket{m_1}\!\!\bra{m_1}\!-\!\ket{m_2}\!\!\bra{m_2}$, and we neglected the terms proportional to the identity.

Considering the validity of the equivalence principle between inertial and gravitational mass, and representing the gravitational interaction with the Newtonian potential, the Hamiltonian of $N$ mixed particles interacting gravitationally becomes
\begin{eqnarray}
\label{Hamiltoniana_large}
H^{(N)}=\sum_i \omega_i \sigma_i^{(z)}+ \frac{1}{2}\sum_{i,j}\Omega_{i,j} \sigma_i^{(z)}\cdot\sigma_j^{(z)} \, ,
\end{eqnarray}
where  $\omega_i=\omega_0+\sum_{j}g_{i,j}(m_1^2-m_2^2)$ with  $\omega_0=\frac{c^2}{4 E}(m_1^2-m_2^2)$, \mbox{$\Omega_{i,j}=g_{i,j}(m_1-m_2)^2$} with $g_{i,j}=\frac{G}{4d_{i,j}}$, $\Omega_{i,i}=0$ and $d_{i,j}$ relative distance between the $i$-th and the $j$-th fields that, for sake of simplicity is assumed constant during the evolution.

Assuming that at time $t\!=\!0$ the system is described by a fully separable state and that the first $M$ particles are created in the state $\ket{n_A}$ and the rest is in the state $\ket{n_B}$, then the initial state is \mbox{$\ket{\psi^{(N)}(0)}=\bigotimes_{\alpha=1}^M\ket{n_A}_\alpha\bigotimes_{\beta=M+1}^{N-M}\ket{n_B}_\beta$}.
For $t>0$, since the system is closed, one has the pure state \mbox{$\ket{\psi^{(N)}(t)}=U(t)\ket{\psi^{(N)}(0)}$}, where the unitary evolution operator is $U(t)=\exp(-\imath t H^{(N)})$.
The reduced density matrix on the selected $k$-th particle  is
\begin{eqnarray}
\label{reduced_density_general}\nonumber
\!\rho_{k}(t)\!&=&\!\frac{1}{2}\!\left( \! \mathbb{1}\!+\!
\sum_{\alpha} \! \bra{\psi^{(N)}(0)}\! U^\dagger\!(t) \sigma^\alpha_k U\!(t) \!\ket{\psi^{(N)}(0)}  \sigma^\alpha_k \right)\!\!
\\\nonumber \!&=&
\!\frac{1}{2}\!\left(
\begin{array}{cc}
1 +\zeta_k \cos(2 \theta) & \zeta_k e^{-\imath \phi} \sin(2 \theta ) a_k^*(t) \\
\zeta_k e^{\imath \phi} \sin(2 \theta ) a_k(t) & 1 -\zeta_k \cos(2 \theta)
\end{array}
\right),\!\!\!\!\!\!\!
\end{eqnarray}
where the index $\alpha$ runs over the ensemble $\{x,y,z\}$, the function $\zeta_k$ is equal to $+1$ for $k\le M$, and to $-1$ for $k > M$, and $a_k(t)$ is given by
\begin{equation}
\label{definition_of_a}
 a_k(t)\! =\! e^{\imath 2 \omega_k t }\!\prod_{j=1}^N\!(\cos(2 \Omega_{k,j}t)\!+\!\imath \zeta_k\cos(2\theta)\sin(2 \Omega_{k,j}t)).\!\!\!\!
\end{equation}

Considering two copies of the $N$-particle system such that in the first copy one has $M=N$ and in the second one has $M=0$, the oscillation probabilities (obtained considering the average over all the elements of the system) are
\begin{eqnarray}
\label{probability_interacting_many}
P_{n_A \rightarrow n_B}(t) \! &\!=\!& \frac{1}{2} \sin^2(2\theta)\left(1 - \frac{1}{N}\!\sum_{k=1}^N  \mathrm{Re}(a_k^{(A)}(t))\right) \,;
\nonumber \\
P_{n_B \rightarrow n_A}(t) \! &\!=\!& \frac{1}{2} \sin^2(2\theta)\left(1 - \frac{1}{N}\!\sum_{k=1}^N  \mathrm{Re}(a_k^{(B)}(t))\right)\,,
\end{eqnarray}
where $\mathrm{Re}(a_k^{(A)}(t))$ is the real part of $a_k(t)$ when $M\!=\!N$, and similar for $\mathrm{Re}(a_k^{(B)}(t))$.
Note that the $CP$-symmetry is preserved, indeed \mbox{$\Delta_{CP}(t)\!\! =\! P_{n_A \rightarrow n_B}(t)\! -\! P_{\overline{n}_A \rightarrow \overline{n}_B}(t)\!=\!\! 0$}, where $\overline{n}_\sigma$ ($\sigma\! =\! A, B$) is the antiparticle state.
On the contrary since, $a_k^{(A)}(t)\neq a_k^{(B)}(t)$, the $T$-symmetry is violated. In fact we have \mbox{$\Delta_T(t) = P_{n_A \rightarrow n_B}(t) - P_{n_B \rightarrow n_A}(t)\neq 0$.}
Assuming $\Omega_{i,j} t\ll 1$ the expression of $a_k(t)$ simplifies and becomes equal to $a_k(t) \simeq e^{\imath \omega_k t } \left(1 \pm 2 \imath \cos(2 \theta) \sum_{j=1}^N \Omega_{k,j}t \right),$ where the sign $+$ is for $a_k^{(A)}(t)$ and the sign $-$ is for   $a_k^{(B)}(t)$. Then $\Delta_T$ becomes
\begin{eqnarray}
\label{probability_interacting_many_2}
 \Delta_T(t) =
\sin^2(2 \theta) \cos(2\theta) \frac{2 t}{N}\!\! \sum_{k,j=1}^N \sin(2 \omega_k t) \Omega_{k,j}\,.
\end{eqnarray}
Being $\Delta_{CP} \neq \Delta_T$, also the $CPT$ symmetry is violated.
Denoting with $F=\frac{1}{N}\sum_{k=1}^N f_k$ the average of $f_k$, which are defined as $f_k=\frac{\sin(2 \omega_k t)}{N}\sum_j \Omega_{k,j}$,
one can express $\Delta T$ in terms of average values of relative distances among the particles in the system as
\begin{equation}\label{GenericDeltaT}
\Delta_T = \sin^2(2 \theta) \cos(2\theta) 2 N t F\,.
\end{equation}
This relation shows the explicit dependence of $\Delta_T$ on the number of particles of the system.
Similar results are obtained for all the configurations in which, $n_A - n_B \sim N$ at $t = 0$.
If $n_A \sim n_B  $ at $t = 0$, one has  $\Delta_T \propto \sqrt{N}$.

 \section{Trapped ions as gravitationally interacting particles}

Let us now consider cooled ions held in a segmented ion trap and exposed to a magnetic field gradient in order to realize effective spin $1/2$ models.
The effective spin-spin interactions induced by the magnetic field are of Ising type and can be adjusted by tailoring the axial trapping potential.
In particular, if the ions are sufficiently cold, such that the ion motion can be neglected, the effective system of $N$ spins is described by the Ising Hamiltonian~\cite{Trap6}

\begin{eqnarray}\label{Ising}
\bar{H}_{\text {Ising }}^{(z)} =\frac{\hbar}{2} \sum_{i=1}^{N} \overline{\omega}_{i} \sigma_{i}^{z}
-\frac{\hbar}{2} \sum_{i, j} J_{i j} \sigma_{i}^{z} \sigma_{j}^{z}
\end{eqnarray}
where $\overline{\omega}_{i}$ are the resonance frequencies of the atomic spins, depending on the external magnetic field $B\left(x_{0, j}\right)$ at the equilibrium position of the ion $x_{0, j}$.
The spin-spin couplings $J_{i j}$ depend on the trapping potential and on the spatial derivative of the spin resonance frequency, that is determined by the magnetic field gradient.
They are given by
\begin{equation}
J_{i j}=\left.\left.\frac{\hbar}{2} \frac{\partial \overline{\omega}_{i}}{\partial x_{i}}\right|_{x_{0, i}} \frac{\partial \overline{\omega}_{j}}{\partial x_{j}}\right|_{x_{0, j}}\left(A^{-1}\right)_{i j}
\end{equation}
where $A$ is the Hessian matrix of the potential energy function $V\left(x_{1}, \cdots x_{N}\right)$
\begin{equation}
A_{i j}=\left.\frac{\partial^{2} V\left(x_{1}, \cdots x_{N}\right)}{\partial x_{i} \partial x_{j}}\right|_{x_{\ell}=x_{0, \ell}, \forall \ell}
\end{equation}
that confines the ions in the position $x_{j}$.
In addition, the magnetic gradient allows for the addressing of individual spins with a microwave field, which can be used to manipulate the spin dynamics.

By comparing the Hamiltonians (\ref{Hamiltoniana_large}) and (\ref{Ising}), we see that they are formally identical, with the obvious correspondence $\omega_i \leftrightarrow\frac{\overline{\omega}_{i}}{2} \ \ \ , \ \ J_{i j} \leftrightarrow - \Omega_{i,j}$.
Hence the resonance frequencies of the atomic spins $\overline{\omega}_{i}$ play the role of ${\omega}_{i}$ in Eq.(\ref{Hamiltoniana_large}), apart from a factor $1/2$, and the spin-spin couplings $J_{i j}$ correspond to the gravitational couplings $\Omega_{i,j}$ except for a sign.
Thus the results presented for mixed particle systems can be reproduced with trapped ions.

It is clear, however, that the analogy fails when the antiparticles are involved.
This is because one does not have the $CP$ conjugate of the system at his disposal in the case of trapped ions.
If one could hypothetically reproduce $CP$ conjugate of the system, and the only interaction within the system were the one of Eq. \eqref{Ising}, all the results, including the $CP$ symmetry would be recovered.
In any case, since the oscillation probability are not dependent on the $CP$ violating Majorana phase, there is no reason to believe that the $CP$ conjugate oscillations differ from Eq. \eqref{probability_interacting_many}. Then we can infer that the trapped ion system is also $CP$ invariant. In this sense, we can assume that $\Delta_{CP} = 0$.
Since the $T$ violation can be tested in this setup, and in the neutrino mixing case it induces a $CPT$ symmetry violation,
trapped ions can be used to indirectly test the $CPT$ violation predicted for neutrinos gravitationally interacting.

We now proceed with a numerical analysis, considering different potential shapes, and we
analyze the oscillation formulae and the $T$ violation resulting  for the cases
in which the trap is loaded with $4$ or $6$ ions. The initial state of the ion ensemble is assumed to be factorizable as the direct product of single ion states. We consider two possible initial single ion states, written as superposition of spin up $\ket{\uparrow} $ and down $\ket{\downarrow}$:
\begin{eqnarray}
\label{ion_state}
\ket{ A} & = & \cos(\theta) \ket{\uparrow}+ e^{\imath \phi } \sin(\theta) \ket{\downarrow} \ \nonumber \, \\
\ket{ B} & = & - e^{-\imath \phi } \sin(\theta) \ket{\uparrow}+ \cos(\theta) \ket{\downarrow} \, .
\end{eqnarray}
Notice that both the states of Eq. \eqref{ion_state} are pure.

We start by studying three different wells loaded with 4 strongly coupled ions. These simulate a non--uniform distribution of neutrinos in space.
To investigate a realistic experimental set-up of this configuration, we use the following values of resonance frequencies of the atomic spins splitting: $\Delta \omega_{i} / 2 \pi(\mathrm{MHz})\begin{array}{llll} = -30.7; & -2.3; & 2.4; & 30.9
\end{array}$, and the following values of the coupling constant $J_{i j}(\mathrm{~Hz})$: \mbox{$ J_{1,2} = 2.1; $} $J_{1,3} =1.8$; $ J_{1,4} =0.4; $ $J_{2,3} =123.8 $, with $J_{3,4} = J_{2,4}$~\cite{Trap-d}.

\begin{figure}[t]
\begin{picture}(300,180)(0,0)
\put(10,20){\resizebox{8.0 cm}{!}{\includegraphics{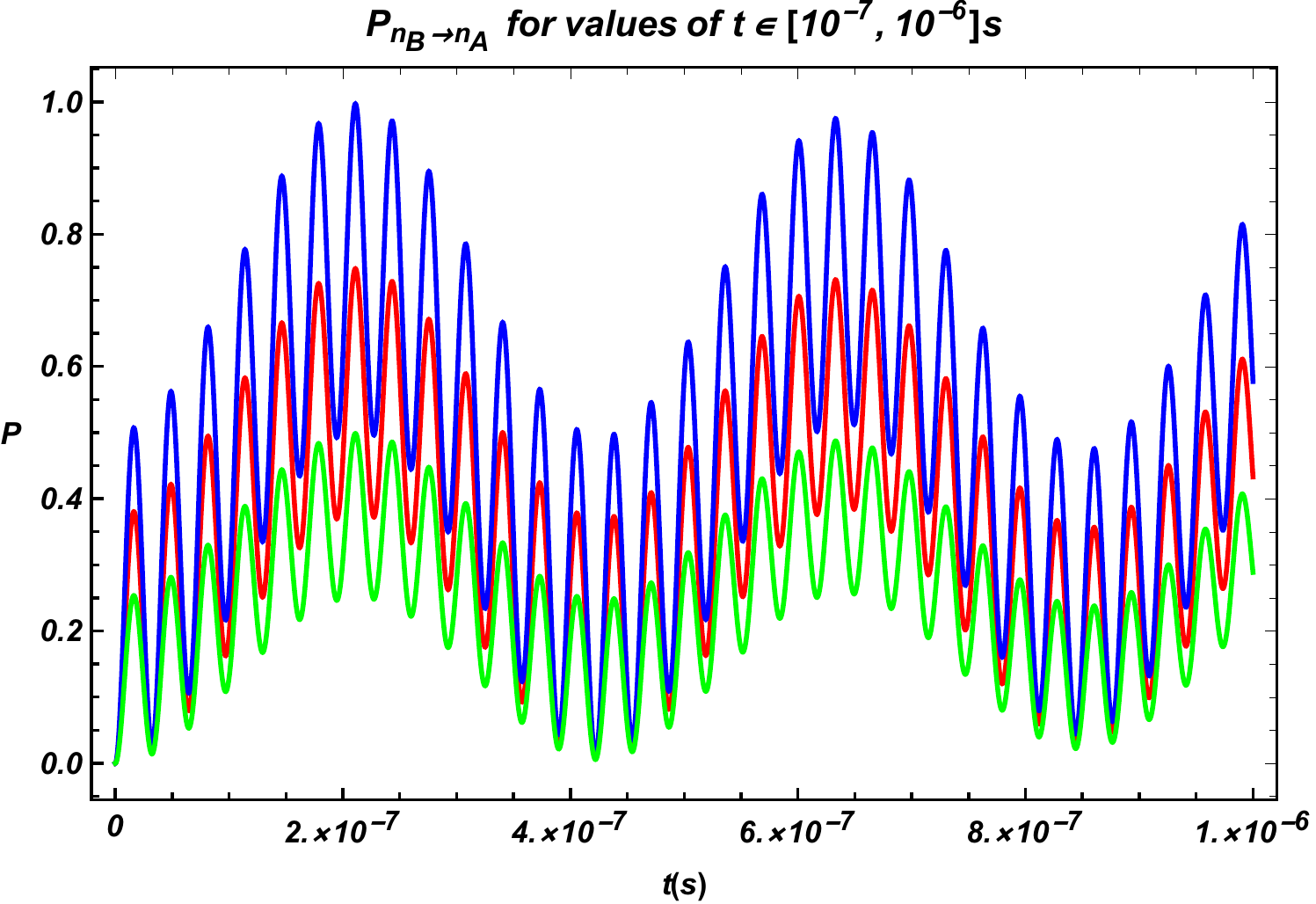}}}
\end{picture}
\caption{\em (Color online) Plots of $  P_{ {B} \rightarrow  {A}}(t)   $, for $\theta=\pi/3$ (the red line), $\theta=\pi/4$ (the blue line), $\theta=\pi/8$  (the green line).  In the plots, we consider $n=4$ trapped ions, and the following values of the  resonance frequencies of the atomic spins splitting $\Delta \omega_{i} / 2 \pi(\mathrm{MHz})\begin{array}{llll} = -30.7; & -2.3; & 2.4; & 30.9
\end{array}$, and  of the coupling constant  $J_{i j}(\mathrm{~Hz})$: $ J_{1,2} = 2.1; $ $J_{1,3} =1.8$; $ J_{1,4} =0.4; $ $J_{2,3} =123.8
 $, with $J_{3,4} = J_{2,4}$.}
\label{Osc1}
\end{figure}

\begin{figure}[t]
\includegraphics[width=\columnwidth]{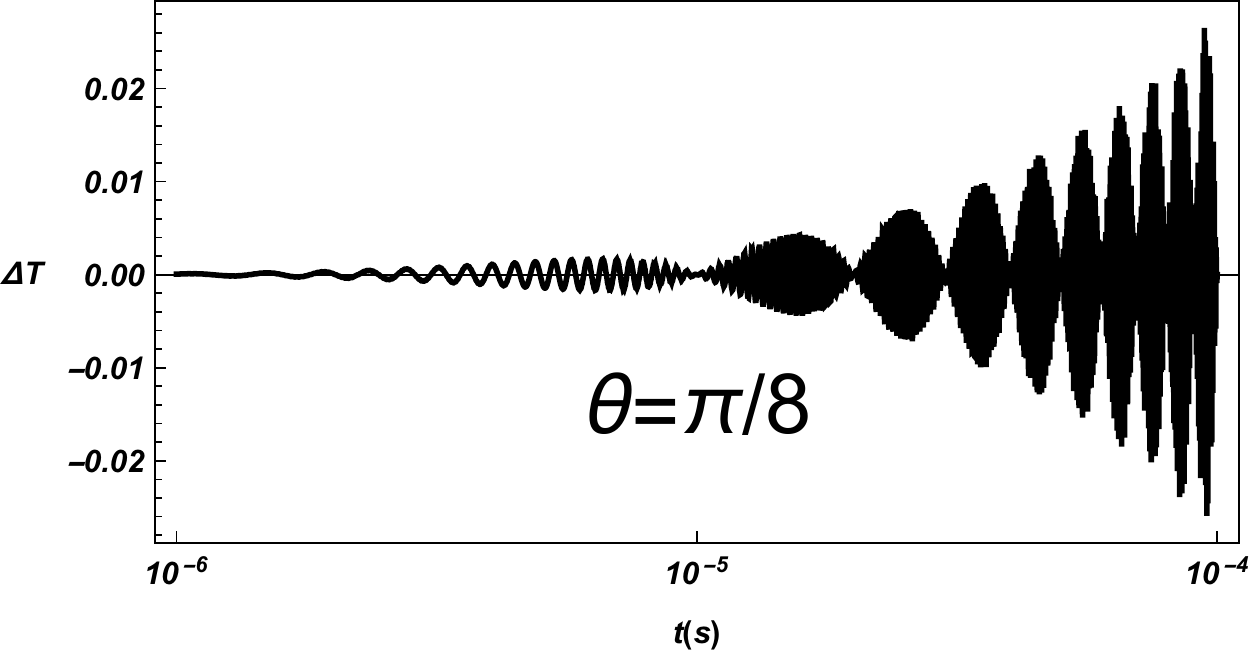}
\includegraphics[width=\columnwidth]{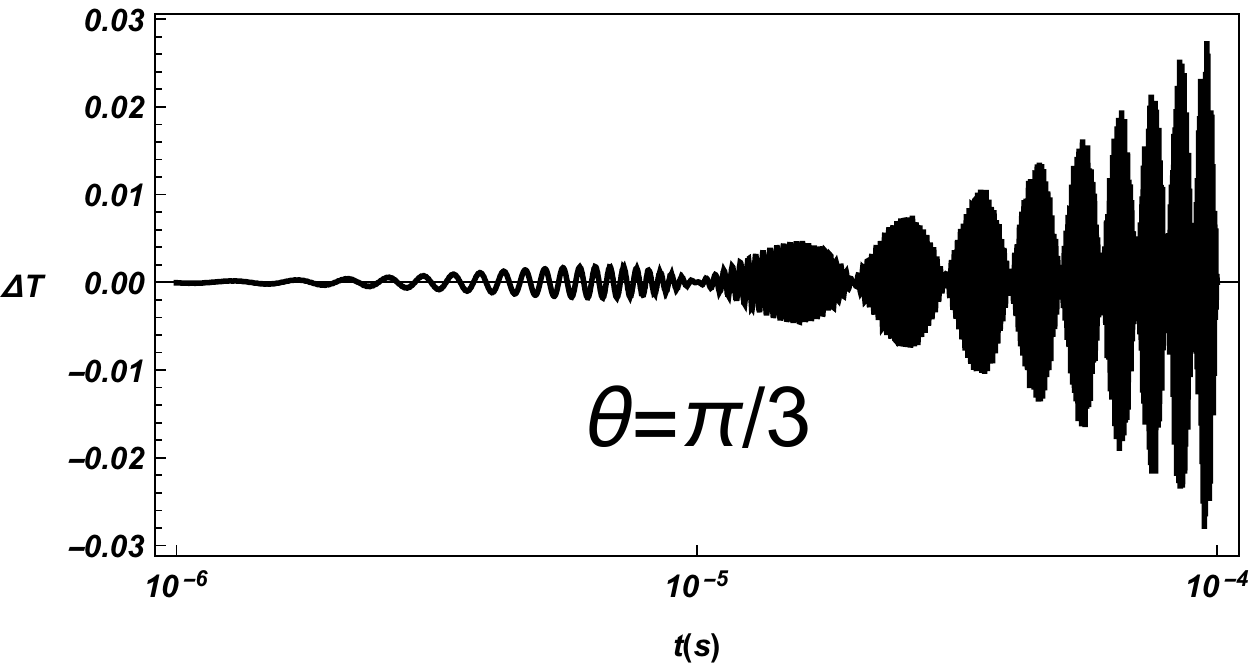}
\includegraphics[width=\columnwidth]{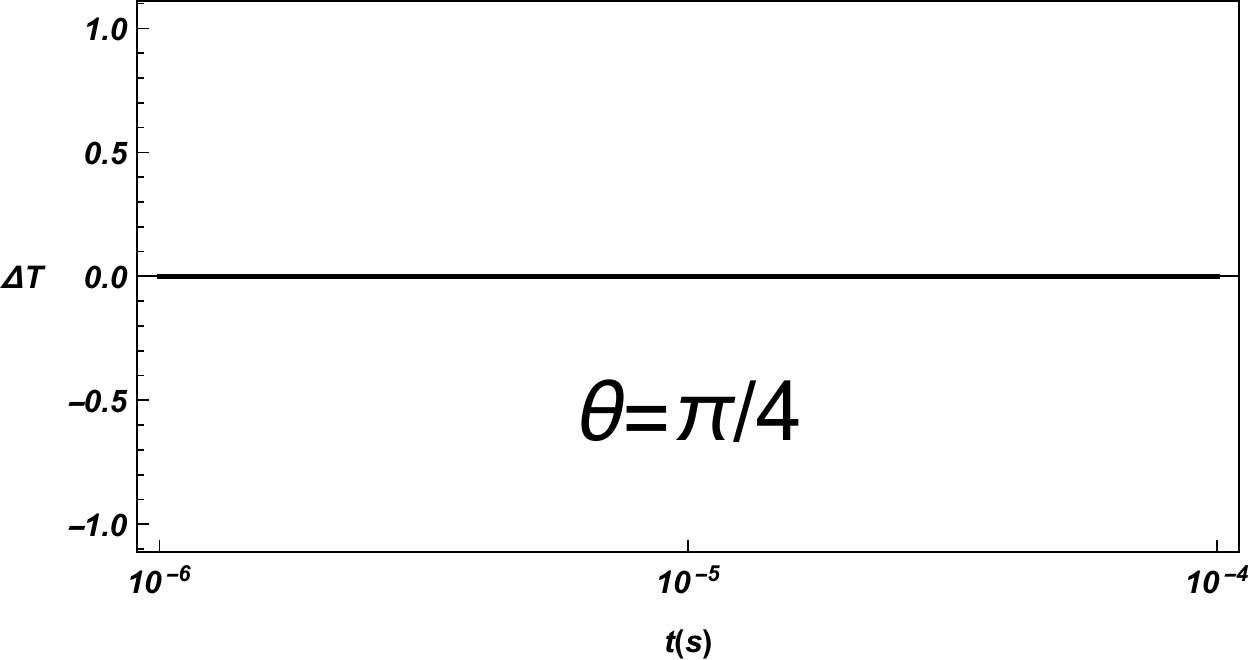}
\caption{Plots of $\Delta_{T}(t)$ for different choices of $\theta$ for a set-up with $n=4$ trapped ions, and the same parameters as figure (1). Here we use a different time scale in order to better highlight the $T$-Violation. }
\label{pdf_1}
\end{figure}

Plots of the transition probability  $P_{ {B} \rightarrow  {A}} $  as a function of time are shown in Figs.(\ref{Osc1}) and the results for the violation of the $T$-symmetry $\Delta_{T} $ as a function of time for such a system are reported in Fig.~\ref{pdf_1}. For the oscillation probabilities (Fig. \ref{Osc1}) we use the time range $[10^{-7}- 10^{-6} ]\ \mathrm{s}$, while the $T$-asymmetries are plotted in the range $[10^{-6}-10^{-4}] \ \mathrm{s}$ because the magnitude of the oscillations is larger here, and therefore they are more visible.

The plots show that, apart from specific singular values, e.g. $\theta=\pi/4$, the violation of the $T$-symmetry, and then of the $CPT$ symmetry, is large enough to be detected with the current technologies. Notice that the time intervals considered ($t \in [10^{-7}- 10^{-4]}$ seconds), are well below the coherence time usually characterizing the experiments with trapped ions.

\begin{figure}[t]
\begin{picture}(300,180)(0,0)
\put(10,20){\resizebox{8.0 cm}{!}{\includegraphics{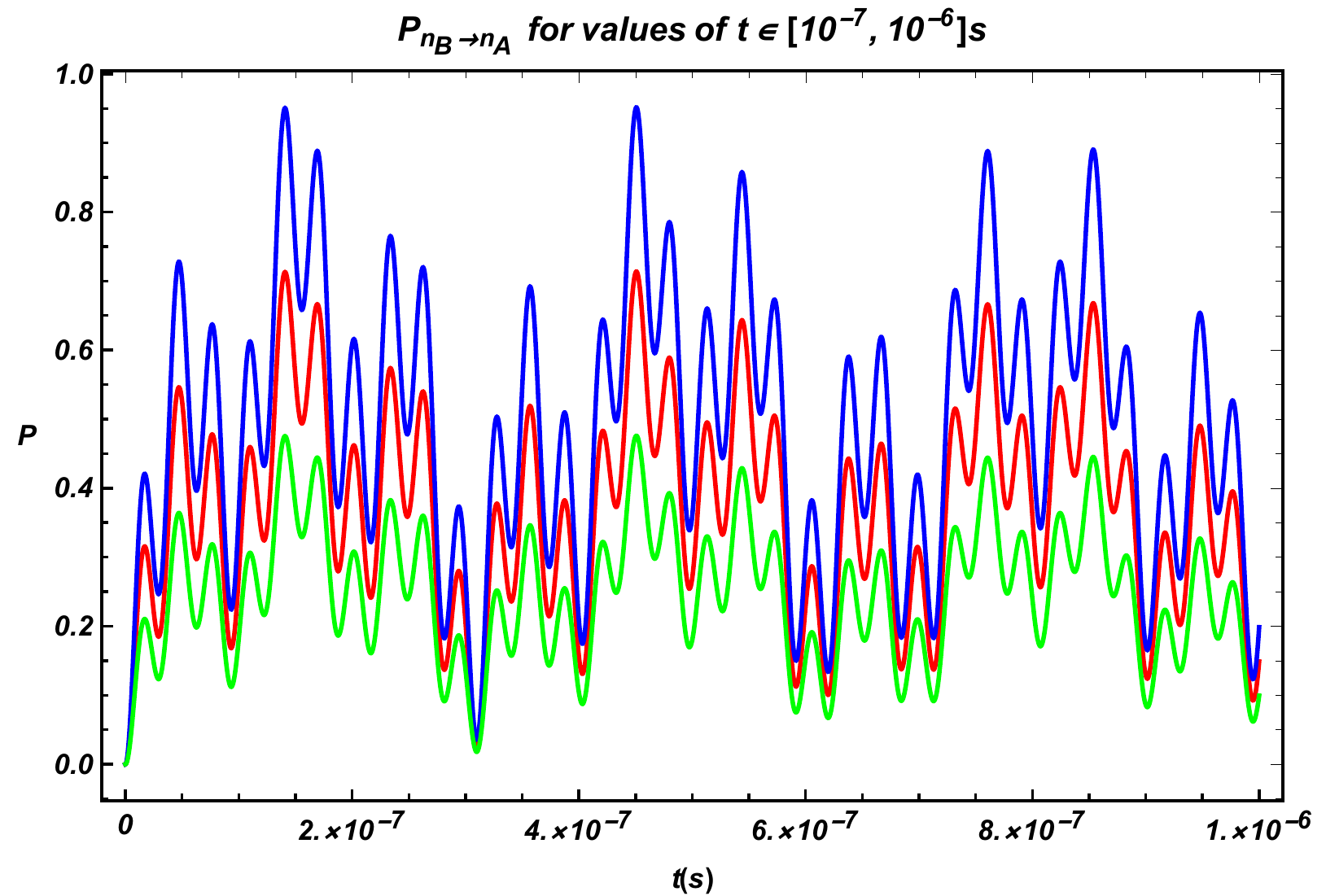}}}
\end{picture}
\caption{\em (Color online) Plots of $  P_{n_{B} \rightarrow n_{A}}(t)   $, for $\theta=\pi/3$ (the red line), $\theta=\pi/4$ (the blue line), $\theta=\pi/8$  (the green line).  In the plots, we consider $n=6$ trapped ions, and the following values of the  resonance frequencies of the atomic spins splitting $\Delta \omega_{i} / 2 \pi(\mathrm{MHz})\begin{array}{llllll} = -32.01; & -9.9; & -3.0; & 3.2; & 10.0; & 32.3\end{array}$, and  of the coupling constant  $J_{i j}(\mathrm{~Hz})$: $ J_{1,2} = 27.9; $ $J_{1,3} = 19.5$; $ J_{1,4} =16.7; $ $J_{1,5} =16.7;
 $ $J_{1,6} =1.4; $ $J_{2,3} =411.8; $   $J_{2,4} = 319.7; $    $J_{2,5} = 300.3; $  $J_{2,6} = 16.5; $
 $J_{3,4} =348.3; $ $J_{3,5} =319.2; $   $J_{3,6} = 16.4; $    $J_{4,5} = 410.9; $  $J_{4,6} = 19.1; $ $J_{5,6} = 27.3, $
  and $J_{i,j} = J_{j,i}$.}
\label{Osc2}
\end{figure}

\begin{figure}[t]
\includegraphics[width=\columnwidth]{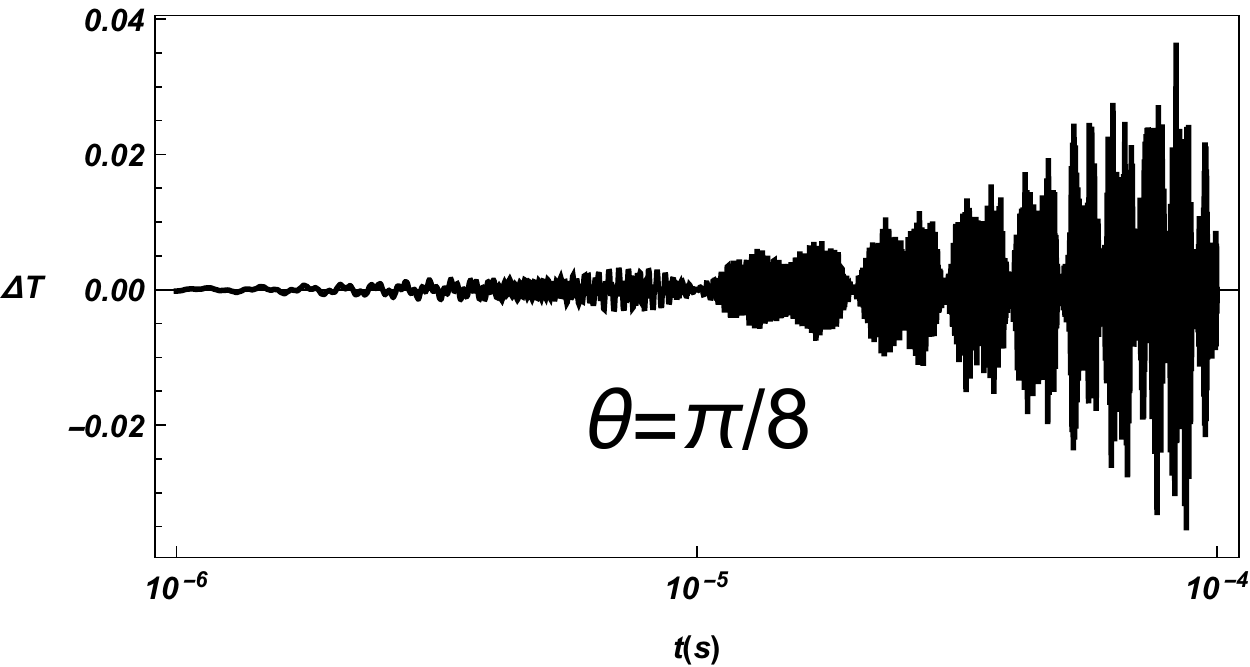}
\includegraphics[width=\columnwidth]{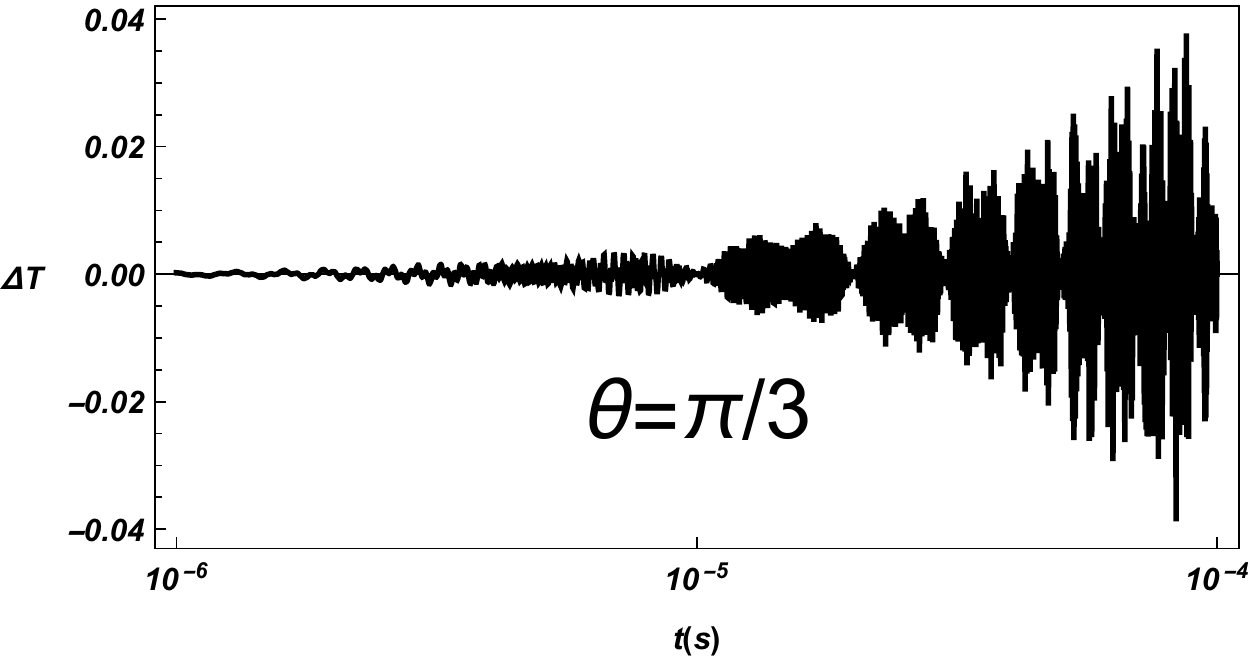}
\includegraphics[width=\columnwidth]{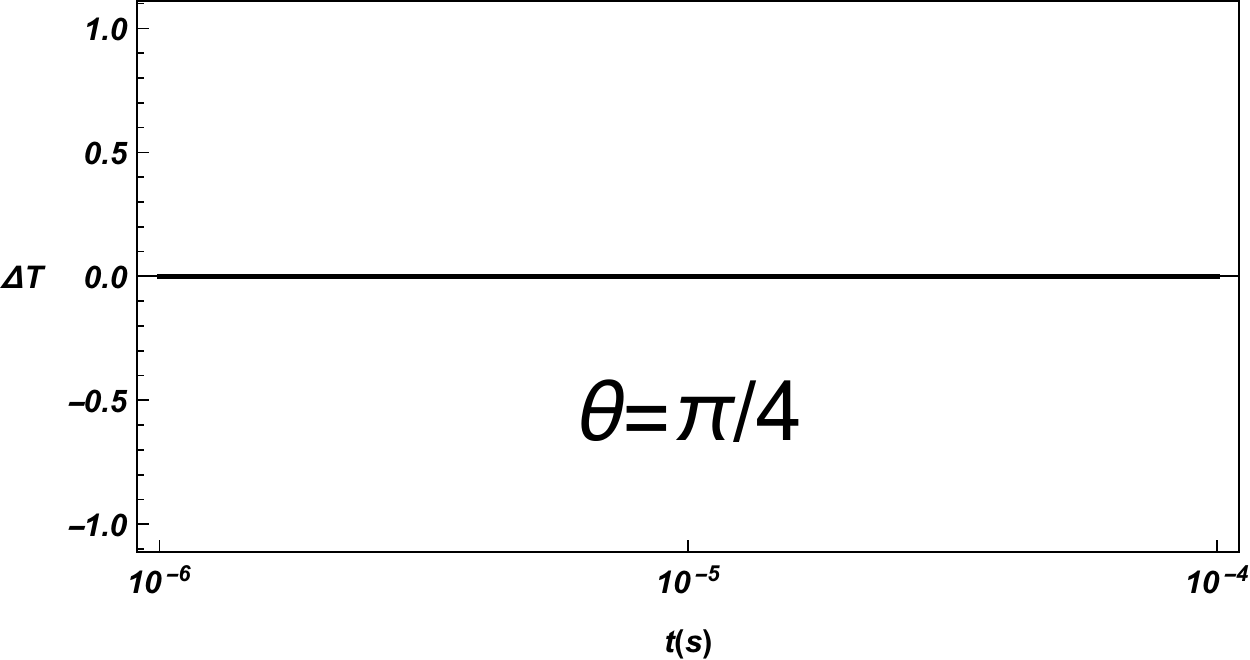}
\caption{Plots of $\Delta_{T}(t)$ for different choices of $\theta$ for a set-up with $n=6$ trapped ions. The same parameters as figure (3) are used. We used a different timescale to better highlight the $T$-violation.}
\label{pdf_2}
\end{figure}

With today's technologies of trapped ions we can also simulate larger systems. However an increment of the number of ions implies a reduction of the coherence time. Keeping the total number of ions sufficiently small, like $n=6$, the coherence time is still quite large.
Then, we take into account a system made by $n=6$ ions trapped in a single well with $\Delta \omega_{i} / 2 \pi(\mathrm{MHz})\begin{array}{llllll} = -32.01; & -9.9; & -3.0; & 3.2; & 10.0; & 32.3\end{array}$, and   $J_{i j}(\mathrm{~Hz})$: $ J_{1,2} = 27.9; $ $J_{1,3} = 19.5$; $ J_{1,4} =16.7; $ $J_{1,5} =16.7;
 $ $J_{1,6} =1.4; $ $J_{2,3} =411.8; $   $J_{2,4} = 319.7; $    $J_{2,5} = 300.3; $  $J_{2,6} = 16.5; $ and
$J_{3,4} =348.3; $ $J_{3,5} =319.2; $   $J_{3,6} = 16.4; $    $J_{4,5} = 410.9; $  $J_{4,6} = 19.1; $ $J_{5,6} = 27.3, $
and $J_{i,j} = J_{j,i}$~\cite{Trap-d}.
Considering the same initial states of the previous case, we have the oscillation probabilities shown in fig. \ref{Osc2} and the $T$-symmetry violation shown in fig.~\ref{pdf_2}.
Also here we use different scales for the oscillation probabilities (Fig. \ref{Osc1})  $[10^{-7}-10^{-6}] \ \mathrm{s}$, and the $T$-asymmetries  $[10^{-6}-10^{-4}] \ \mathrm{s}$, to better highlight the latter.

It should be noted that the general prediction of Eq. \eqref{GenericDeltaT} of the dependence of the $T$ asymmetry on the total particle number $N$ is confirmed by the plots shown above.
The $6$ ions asymmetry of fig.~\ref{pdf_2} (given the different couplings) at $t_1 = 10^{-6} s$ is slightly bigger than the $4$ ions asymmetry of fig.~\ref{pdf_1} at the same time $t_1$ and, due to Eq. \eqref{GenericDeltaT}, we expect that a larger number of ions implies a larger $T$ asymmetry.
In any case, a limited number of ions is to be preferred, since despite a larger signal, the coherence time is reduced when a large number of ions is considered.
The plots obtained show that the oscillation formulae and the $CPT$ violation predicted for gravitationally interacting neutrinos
can be tested with ions trapped in many  potentials.

\section{Conclusions}

The  gravitational self-interaction in particle mixing systems leads to new oscillation formulae  and to the
$CPT$ violation. The corrections to the flavor transitions and the expected $CPT$ symmetry breaking are negligible in the present epoch and therefore they are very hard to be revealed.
However, they are proportional to the number of elements of the system and to its density, so that
these effects    could affect extremely dense systems with a large number of particles, such as some galactic objects and the primordial stages of the universe.
Moreover,  the mechanism leading to $CPT$ violation is not only limited to gravitational interaction. Therefore, the possibility   of testing these processes is of great importance for the understanding of fundamental physics.

In our work, we have shown  the analogies between the effective system of $N$ spins described by the Ising Hamiltonian
and the phenomenon of the gravitational self interaction in mixed particle systems.
 We have shown that cooled ions held in many potential traps
  can simulate the mutual interaction in mixed neutrino system and allow to reveal the expected new oscillation formulae and the $CPT$ violation for this phenomenon.
We considered  the case with $ N= 4$ ions and coupling in three independent wells, and the case of   $6$ ions confined in a single strongly an-harmonic well. We have shown that the $T$ violation for trapped ions, corresponding to the $CPT$ violation for flavor mixed particles, grows with the size of the system $N$.
Our numerical results show that such violations, together with the new oscillation formulae are all detectable in the present experiments on ion traps. Finally, note that in our approach we have chosen the couplings $J_{ij}$ and the frequencies $\bar{\omega}_i$ in order to maximize the $T$--violation. By lowering the interaction, and therefore the couplings and the frequencies, it is possible to slow down and highlight the oscillations. The potentials can always be tuned according to whether one is interested in highlighting the $T$ asymmetry or the oscillation formulae.

\section*{Acknowledgements}

A.C. and A.Q thank  partial financial support from MIUR and INFN. A.C. also thanks the COST Action CA1511 Cosmology and Astrophysics Network for Theoretical Advances and Training Actions (CANTATA).
SMG acknowledge support from the European Regional Development Fund for the Competitiveness and Cohesion Operational Programme (KK.01.1.1.06--RBI TWIN SIN) and from the Croatian Science Fund Projects No. IP-2016--6--3347 and No. IP-2019--4--3321.
SMG also acknowledge the QuantiXLie Center of Excellence, a project co--financed by the Croatian Government and European Union through the European Regional Development Fund--the Competitiveness and Cohesion Operational Programme (Grant KK.01.1.1.01.0004).


\begin{thebibliography}{99}

\bibitem{Wu1957}
 C.~S.~Wu, E.~Ambler, R.~W.~Hayward, D.~D.~Hoppes, and R.~P.~Hudson, Phys. Rev. \textbf{105}, 1413 (1957).

\bibitem{Giunti2007}
C.~Giunti and C.~W.~Kim, \textit{Fundamentals of Neutrino Physics and Astrophysics} (Oxford University Press, 2007).

\bibitem{Griffiths2008}
D.~J.~Griffiths, \textit{Introduction to Elementary Particles} (Wiley-VCH, 2008).

\bibitem{Raffelt1996}
G.~G.~Raffelt, \textit{Stars as Laboratories for Fundamental Physics} (University of Chicago Press, 1996).

\bibitem{Sikivie2011}
P.~Sikivie, Phys. Lett. B \textbf{695}, 22 (2011).

\bibitem{Capolupo2019_1}
A.~Capolupo, I.~De~Martino, G.~Lambiase and A.~Stabile, Phys. Lett. B \textbf{790}, 427 (2019).

\bibitem{Pham2015}
T.~N.~Pham, Phys. Rev. D \textbf{92}, 054021 (2015).

\bibitem{Christenson1964}
J.~H.~Christenson, J.~W.~Cronin, V.~L.~Fitch and R.~Turlay, Phys. Rev. Lett. \textbf{13}, 138 (1964).

\bibitem{Abashian2001}
A.~Abashian et al., Phys. Rev. Lett. \textbf{86}, 2509 (2001).

\bibitem{Bilenky1978}
S.~M.~Bilenky and B.~Pontecorvo, Phys. Rep. \textbf{41}, 225 (1978).

\bibitem{Gonzalez2008}
M.~C.~Gonzalez-Garcia and M.~Maltoni, Phys. Rep. \textbf{460}, 1 (2008).

\bibitem{Phillips2016}
D.~G.~Phillips II et al., Phys. Rep. \textbf{612}, 1 (2016).

\bibitem{Nakamura2010}
K.~Nakamura et al. [Particle Data Group], J. Phys. G \textbf{37}, 075021 (2010).




 \bibitem{Bose2017}
 S.~Bose, A.~Mazumdar, G.~W.~Morley, H.~Ulbricht, M.~Toro\v{s}, M.~Paternostro, A.~A.~Geraci, P.~F.~Barker, M.~S.~Kim, and G.~Milburn, Phys. Rev. Lett. \textbf{119}, 240401 (2017).

 \bibitem{Marletto2017}
 C.~Marletto and V.~Vedral, Phys. Rev. Lett. \textbf{119}, 240402 (2017).

\bibitem{Marletto2018}
C.~Marletto, V.~Vedral, and D.~Deutsch, New J. Phys. \textbf{20}, 083011 (2018).

\bibitem{Kyrylo2019}
K. Simonov, A. Capolupo and S. M. Giampaolo, Eur. Phys. J. C \textbf{79} 11, 902 (2019).





\bibitem{Ellis1984}
J.~Ellis, J.~S.~Hagelin, D.~V.~Nanopoulos, and M.~Srednicki, Nucl. Phys. B \textbf{241}, 381 (1984).

\bibitem{Amelino-Camelia2000}
G.~Amelino-Camelia, in \textit{Towards Quantum Gravity}, edited by J.~Kowalski-Glikman (Springer, 2000), pp.~1--49.

\bibitem{Rovelli2004}
C.~Rovelli, \textit{Quantum Gravity} (Cambridge University Press, 2004).

\bibitem{Lisi2000}
E.~Lisi, A.~Marrone, and D.~Montanino, Phys. Rev. Lett. \textbf{85}, 1166 (2000).

\bibitem{Gago2001}
A.~M.~Gago, E.~M.~Santos, W.~J.~C.~Teves, and R.~Zukanovich~Funchal, Phys.~Rev.~D \textbf{63}, 073001 (2001).

\bibitem{Guzzo2016}
M.~M.~Guzzo, P.~C.~de~Holanda, and R.~L.~N.~Oliveira, Nucl. Phys. B \textbf{908}, 408 (2016).

\bibitem{Benatti2000}
F.~Benatti and R.~Floreanini, J. High Energ. Phys. \textbf{02}, 032 (2000).

\bibitem{Benatti2001}
F.~Benatti and R.~Floreanini, Phys. Rev. D \textbf{64}, 085015 (2001).

\bibitem{Capolupo2018}
A.~Capolupo, S.~M.~Giampaolo, B.~C.~Hiesmayr, and G.~Vitiello, Phys. Lett. B \textbf{780}, 216 (2018).

\bibitem{Capolupo2019}
A.~Capolupo, S.~M.~Giampaolo, and G.~Lambiase, Phys. Lett. B \textbf{792}, 298 (2019);

\bibitem{Buoninfante2020}
L. Buoninfante, A. Capolupo, S. M. Giampaolo, G. Lambiase, Eur.Phys. J. C \textbf{80} 11, 1009 (2020).

\bibitem{CapolupoCurv2020}
A. Capolupo, G. Lambiase, A. Quaranta Phys. Rev. D \textbf{101}, 095022 (2020).


\bibitem{Trap-a}
M. H. Abobeih, J. Cramer, M. A. Bakker, N. Kalb, M.
Markham, D. J. Twitchen, T. H. Taminiau, Nature comm. \textbf{9}
2552 (2018)

\bibitem{Trap-b}
N. Friis, O. Marty, C. Maier, C. Hempel, M. Holz\"{a}pfel, P. Jurcevic, M. Plenio, M. Huber, and C. Roos, Physical Review X. \textbf{8} 021012, (2018)

\bibitem{Trap1}
S. A. Schulz, U. Poschinger, F. Ziesel, and F.Schmidt-Kaler,
New J. Phys. \textbf{10}, 045007 (2008).

\bibitem{Trap-c}
D. Kielpinski, C. Monroe, and D. J.  Wineland Nature. \textbf{417}, 709 (2002)

\bibitem{Trap-d}
S. Zippilli, M. Johanning, S. M. Giampaolo, Ch. Wunderlich, and F. Illuminati
Phys. Rev. A \textbf{89}, 042308 (2014)

\bibitem{Trap2}
D. Kaufmann, T. Collath, M. T. Baig, P. Kaufmann, E. Asenwar, M. Johanning, and Ch. Wunderlich, Applied Physics B \textbf{107}, 935 (2012).

\bibitem{Trap3}
F. Mintert and C. Wunderlich, Phys. Rev. Lett. \textbf{87}, 25 (2001).

\bibitem{Trap4}
J. Chiaverini, W. E. Lybarger, Phys. Rev. A \textbf{77}, 022324 2008.

\bibitem{Trap5} A. Khromova, C. Piltz, B. Scharfenberger, T. F. Gloger, M. Johanning, A. F. Var\'on, and C. Wunderlich, Phys. Rev. Lett. \textbf{108},
220502 (2012).

\bibitem{Trap6}
M. Johanning, A. F. Varon, and C. Wunderlich, J. Phys. B: At. Mol. Opt. Phys. \textbf{42}, 1 (2009).




\end{thebibliography}
\end{document}